\documentstyle[twocolumn]{jpsj}
\def\runtitle{Electronic Structure of B-2$p$ State in AlB$_2$ Single Crystal: 
Direct Observation of $p\sigma$ and $p\pi$ Density of States}
\def\runauthor{Jin {\sc Nakamura} et al.}

\title
{
Electronic Structure of B-2$p$ State in AlB$_2$ Single Crystal: \\
Direct Observation of $p\sigma$ and $p\pi$ Density of States
}

\author
{
Jin {\sc Nakamura}\footnote{E-mail: jin@pc.uec.ac.jp}, Masamitsu {\sc Watanabe}$^{1}$, Tamio {\sc Oguchi}$^{2}$, Sin-ya {\sc Nasubida}, Eiki {\sc Kabasawa}, Nobuyoshi {\sc Yamada}, Kazuhiko {\sc Kuroki}, Hisashi {\sc Yamazaki}, Shik {\sc Shin}$^{3}$, Yuji {\sc Umeda}$^{4}$, Shin {\sc Minakawa}$^{4}$, Noriaki {\sc Kimura}$^{4}$ and Haruyoshi {\sc Aoki}$^{4}$
}

\inst
{
Department of Applied Physics and Chemistry, 
The University of Electro-Communications,
Chofu-shi, Tokyo 182-8585\\
$^{1}$RIKEN/Spring-8, 
Kouto 1-1-1, Mikazuki, Sayo, Hyogo 679-5148\\
$^{2}$Department of Quantum Matter, ADSM, 
Hiroshima University, 
Higashihiroshima-shi, Hiroshima 739-8530\\
$^{3}$The Institute for Solid State Physics, 
The University of Tokyo,
Kashiwa-shi, Chiba 277-8581\\
$^{4}$Center for Low Temperature Science, 
Tohoku University,
Sendai-shi, Miyagi 980-8578\\
}

\recdate
{
\today
}

\abst
{
X-ray emission (XES) and absorption (XAS) spectra near the B-$K$ edge were measured on single-crystalline AlB$_2$ compound which is an isostructural diboride of superconducting MgB$_2$.    
The partial density of states (PDOS) of B-2$p\sigma$ and $p\pi$ orbitals were derived from the polarization dependence of XES and XAS spectra.  
There are considerable amounts of PDOS near the Fermi energy in AlB$_2$ similarly to that in MgB$_2$, but there are almost no PDOS in $p\sigma$ orbitals of AlB$_2$ near the Fermi energy, i.e., a pseudo-gap in $p\sigma$ state and a broad metallic state in $p\pi$ state are observed.  
The present result indirectly supports  scenarios that the $p\sigma$ holes play an important role in the occurrence of superconductivity in MgB$_2$.  
The overall features of PDOS were found to be in good agreement with the result of band calculation of AlB$_2$, but a small discrepancy in the Fermi energy is observed, which is attributed to the Al vacancy in the compounds, i.e., the estimated concentration is Al$_{0.93}$B$_2$.   
}

\kword
{
MgB$_2$, AlB$_2$, single crystal, partial density of state, $p\sigma$ and $p\pi$ orbitals, X-ray emission and absorption spectroscopy
}

\begin{document}
\sloppy
\maketitle

Since the discovery of superconductivity in MgB$_2$ with a transition 
temperature, $T_{\rm c}$, of 39 K by Nagamatsu {\it et al.},\cite{Nagamatsu} 
a large number of research studies from experimental\cite{Kurumaev,Callcott,Nakamura} and 
theoretical\cite{Imada,Yamaji,Kortus,Belashchenko,Satta,An,Serrato,Suzuki,Rosner,Medvedeva}  points of view have been performed on this compound and on a series of isostructural diborides.  
In a previous paper,\cite{Nakamura} we reported a large partial density of states (PDOS) of B-2$p$ orbitals near the Fermi energy, $E_{\rm F}$, in MgB$_2$ by soft X-ray emission (XES) and absorption spectroscopies (XAS) near the B-$K$ edge.  
This result is consistent with the results of band calculations, which suggest the holes in $p\sigma$ bands between B-B in a honeycomb plane play important roles in the superconductivity of MgB$_2$.  
Although there are many experimental results that suggest MgB$_2$ is considered as an $s$-type superconductor with a strong electron-phonon coupling, the reason for the high value of $T_{\rm c}$ as a conventional BCS-type superconductor is not clear.  
An efficient step towards understanding the mechanism of superconductivity
in MgB$_2$ is to clarify the difference between this material and
other materials which have the same crystal structure but are
not superconductors. 
An example of such materials is AlB$_2$.  
From a theoretical point of view, 
first principles band calculations reveal that a large difference between MgB$_2$ and AlB$_2$ is that the Fermi level intersects the $2p\sigma$ band in the former, while it does not in the latter, 
suggesting that the $p\sigma$ band plays an important role in the occurrence of superconductivity in MgB$_2$.\cite{Satta,Oguchi}  
However, until now, a direct experimental confirmation that
such band-calculation predictions are indeed correct has not yet been reported.

To clarify this point, here we directly observe the PDOS of
B-2$p\sigma$ and 2$p\pi$ bands in AlB$_2$ by
performing polarization-dependent XES and XAS
on a single crystalline compound.  
The single-crystalline AlB$_2$ samples were prepared by the Al-flux method.  
Mixtures of Al (purity, 4N) and B (purity, 4N5) powders were placed in an Al$_2$O$_3$ crucible and heated in an Ar gas atmosphere up to 1000$^{\circ}$C, and then cooled slowly to 660$^{\circ}$C.
The synthesized AlB$_2$ single crystals were separated from the solidified melts by dissolving the Al flux with sodium hydroxide solution. 
The obtained crystals resemble a hexagonal plate with the edge length (in $ab$-plane) of about 1$\sim$2 mm and with the thickness of about 10 $\mu$m along the $c$-axis.  
Before XES and XAS measurements, the crystal was polished in order to remove Al-flux on the surface of the specimen and mounted on a Au plate with Ag paste.  
The XES measurements were performed at the undulator beamline BL-2C in KEK-PF.\cite{Watanabe}  
The incident photon energy is about 400 eV.   
Emitted photons were detected using the MCP detector combined with the 1200 lines/mm grating.  
The energy resolution of the spectrometer with the slit width of 20 $\mu$m was estimated as about $\Delta E\sim$0.2 eV at the energy of $E$=200 eV.\cite{Harada}  
Polarization dependence of emission spectra on the angle between the $c$-axis and the detector-direction $\theta$ was measured at room temperature (Fig.\ref{FIG1}).  
The XAS measurements were performed at BL-19B in KEK-PF by the total fluorescence yield (TFY) method.  
The energy resolution of the incident photon was about 0.2 eV.  
The geometry of XAS measurement was the same as that of the XES measurement.  

\begin{figure}
\caption{
(a) Experimental setup for soft X-ray emission spectroscopy.  
The hatched area shows the cross section of the hexagonal plate, i.e., $ac$-plane.  
Both the B-2$p\sigma$ and 2$p\pi$ orbitals in the plane are shown.  
In addition to these orbitals, there is another $p\sigma$ component perpendicular to the plane.
(b) In the case of $\theta$=90$^{\circ}$, we observe the emission from both $p\sigma$ and $p\pi$ orbitals with equal weight.
(c)  In the case of $\theta$=0$^{\circ}$, we observe $p\sigma$-emission only.
}\label{FIG1}
\end{figure}

In our experimental geometry, the fluorescence intensity with the incidence angle $\theta$, $I^{fluo}(\theta)$, is expressed using PDOS components parallel to the $p\sigma$ and $p\pi$ orbitals, $I^{fluo}_{p\sigma}$ and $I^{fluo}_{p\pi}$ because of the dipole transition (radiation) from B-2$p$ to 1$s$ states; 
\begin{equation}
\displaystyle 
I^{fluo}(\theta) = [1+\cos^2(\theta)]I^{fluo}_{p\sigma} + \sin^2(\theta)I^{fluo}_{p\pi}.  
\end{equation}
Therefore, an ideal XES spectrum $I^{fluo}(0^\circ)$ contains only the $p\sigma$ component, and $I^{fluo}(90^\circ)$ contains both the $p\sigma$ and $p\pi$ components with equal weight.  

\begin{figure}
\caption{
Theoretical partial density of states of $p_{x}+p_{y} (=2\times p\sigma)$ and $p_{z} (=p\pi)$ orbitals in MgB$_2$ and AlB$_2$ derived from the FLAPW method.\cite{Oguchi}  
}\label{FIG2}
\end{figure}

We first show the theoretical PDOS of $p\sigma$ and $p\pi$ orbitals of MgB$_2$ and AlB$_2$ derived from band calculation (Fig. \ref{FIG2}).\cite{Oguchi}  
It is found that the overall feature of PDOS of MgB$_2$ is almost the same as that of AlB$_2$, i.e., the rigid band model roughly represents these materials.  
This is consistent with the previous XAS and XES results for the polycrystalline samples.\cite{Nakamura}  
To be more precise, the detailed form of PDOS of MgB$_2$ is sharp in comparison with that of AlB$_2$, i.e., the peak of the $p\sigma$ band of MgB$_2$ at $E=-$2 eV is relatively sharp compared with that of AlB$_2$ at $E=-$4.5 eV.  
This is due to a reduction of two-dimensionality of B-$p$ bands in AlB$_2$, which is consistent with the decrease of the lattice-constant ratio from MgB$_2$ ($c/a$=1.14) to AlB$_2$ ($c/a$=1.08).   
As mentioned in the introduction, an important difference between MgB$_2$ and AlB$_2$ predicted theoretically is that the Fermi energy lies within the $p\sigma$ band in the former, but not in the latter.  

\begin{figure}
\caption{
Partial density of states (PDOS) of B-2$p\sigma$ and $p\pi$, 2$\times I^{fluo}_{p\sigma}$ ($\circ$) and $I^{fluo}_{p\pi}$ ($\bullet$), derived from observed $I^{fluo}(20^\circ)$ and $I^{fluo}(45^\circ)$.  
Dotted and solid lines are the theoretical PDOS of $p_{x}+p_{y} (=2\times p\sigma)$ and $p_{z} (=p\pi)$ orbitals derived from band calculation (FLAPW) for AlB$_2$.\cite{Oguchi}
}\label{FIG3}
\end{figure}

We now move on to the experimental results.  
Figure \ref{FIG3} shows the partial density of states (PDOS) of B-2$p\sigma$ and 2$p\pi$, $I^{fluo}_{p\sigma}$ and $I^{fluo}_{p\pi}$, derived from observed $I^{fluo}(20^\circ)$ and $I^{fluo}(45^\circ)$.  
A self-absorption correction was applied to the observed XES spectra before the derivation of $I^{fluo}_{p\sigma}$ and $I^{fluo}_{p\pi}$.    
The area intensities of $I^{fluo}_{p\sigma}$ and $I^{fluo}_{p\pi}$ are normalized to unity in the energy region below 188 eV, and the 2$I^{fluo}_{p\sigma}$ and $I^{fluo}_{p\pi}$ are shown in the figure.  
The value of $E_{\rm F}$ is about 187.5 eV which is about 1.5 eV higher than the value of MgB$_2$ (186 eV), which is in good agreement with the previous report for the polycrystalline AlB$_2$ sample\cite{Nakamura}.  
It is clearly seen that there is almost no PDOS in $p\sigma$ orbitals of AlB$_2$ near $E_{\rm F}$. 
Furthermore, there is a considerable amount of PDOS in $p\pi$ orbitals of AlB$_2$ around $E_{\rm F}$.  

\begin{figure}
\caption{
Partial density of states (PDOS) of B-2$p\sigma$ and $p\pi$, 2$\times I^{abs}_{p\sigma}$ ($\circ$) and $I^{abs}_{p\pi}$ ($\bullet$), derived from observed absorption spectra $I^{abs}(20^\circ)$ and $I^{abs}(70^\circ)$:   
(a) Overall feature of PDOS and (b) comparison with the band calculation.  
Dotted and solid lines are the theoretical PDOS of $p_{x}+p_{y} (=2\times p\sigma)$ and $p_{z} (=p\pi)$ orbitals derived from band calculation (FLAPW) for AlB$_2$.
}\label{FIG4}
\end{figure}

Figure \ref{FIG4} shows the PDOS of B-2$p\sigma$ and $p\pi$, 2$\times I^{abs}_{p\sigma}$ ($\circ$) and $I^{abs}_{p\pi}$ ($\bullet$), derived from observed absorption spectra $I^{abs}(20^\circ)$ and $I^{abs}(70^\circ)$. 
Similarly to the XES spectra, the self absorption correction was applied before their derivation, and the normalized absorption intensity $I^{abs}(\theta)$ is expressed as follows; 
\begin{equation}
\displaystyle 
I^{abs}(\theta) = \sin^2(\theta)I^{abs}_{p\sigma} + \cos^2(\theta)I^{abs}_{p\pi}.  
\end{equation}
In PDOS of $p\pi$ [Fig. \ref{FIG4}(a)], there is large absorption at about 194 eV in contrast to no sharp absorption in $p\sigma$-PDOS, which is assigned to the $p{\pi}$ resonance state on the sample surface.\cite{Callcott}  
In both figures (Figs. \ref{FIG3} and \ref{FIG4}), the theoretical PDOS of AlB$_2$ are shown again by the dotted line ($p\sigma$) and the solid line ($p\pi$).  
It is found that the theoretical PDOS reproduces the observed PDOS well, for both the empty and occupied states.  
We have clearly observed a pseudo-gap of about 3 eV around 187-190 eV in the B-2$p\sigma$ orbital in sharp contrast to the broad metallic state of the B-2$p\pi$ orbital.  
The pseudo-gap is attributed to the bonding and anti-bonding state separation due to the strong covalent nature of $p\sigma$ orbitals.\cite{Oguchi}  
However, there is a small difference in the value of $E_{\rm F}$ between the experimental and theoretical PDOS.  
The band calculation predicts the energy of $p\sigma$-shoulder at about --1.8 eV below the Fermi level.  
However the observed energy of the shoulder (186.3 eV) locates --1.2 eV below the Fermi level (187.5 eV), i.e., the observed Fermi energy $E_{\rm F}$ is 0.6 eV lower than the theoretical prediction.  
The reason for this difference is considered to be the lack of Al atoms from the stoichiometry.  
The observed pseudo-gap in $p\sigma$-PDOS at around the Fermi energy suggests the strong covalent bonding feature of boron forming the 2D honeycomb plane as reported by maximum entropy method (MEM)/Rietvelt analysis.\cite{Nishibori,Lee}  
We considered that the layered B-honeycomb plane is the fundamental structure of AlB$_2$.  
There is a small difference in electronegativity between Al and B atoms, so electrons transfer from Al to B atoms in AlB$_2$ compound.  
Vacancies of Al atoms in AlB$_2$ reduce the number of electrons of B-2$p$ orbitals, thus the Fermi levels shifts down.   
If we assume that the decrease of the states below $E_{\rm F}$ is due to an Al vacancy, the vacancy concentration $x$ of Al$_{1-x}$B$_2$ is estimated to be about 0.07.  
The theoretical band calculation suggests that the heat of formation of AlB$_2$ is lower than that of MgB$_2$, which suggests the compound which has a lower number of electrons of the cation than the stoichiometric AlB$_2$ is more stable.  
The present result is consistent with this prediction.  

To summarize, we have performed direct measurement of partial density of states (PDOS) of B-$2p$ bands in single-crystalline AlB$_2$
using polarization-dependent XES and XAS.  
We have clearly observed a pseudo-gap of about 3 eV in the B-2$p\sigma$ orbital in sharp contrast to the broad metallic state of the B-2$p\pi$ orbital. 
Although the experimentally observed  PDOS is in excellent agreement with
the band calculation results, the Fermi level in the former is found to be
lower by about 0.6 eV than in the latter.
Nevertheless, the Fermi level still lies well above the $p\sigma$ band,
providing a direct confirmation that there are
no $p\sigma$ holes in AlB$_2$. Conversely, considering the
fact that AlB$_2$ is not superconducting, our result
indirectly supports  scenarios that the $p\sigma$ holes play an
important role in the occurrence of superconductivity in MgB$_2$.

We express our thanks to Professor J. Akimitsu of Aoyama Gakuin University for useful discussions.  
We also thank to Mr. T. Takeuchi and Ms. A. Fukushima of ISSP for their technical support in the spectroscopic measurements.  
T.O. gratefully acknowledges the support by a Grant-in-Aid for COE Research 
(No.13CE2002) of the Ministry of Education, Culture, Sports, Science and 
Technology of Japan.  
This experiment was performed under the approval of the Photon Factory 
Advisory Committee (Proposal No. 2001U004).


\begin{thebibliography}{99}
\bibitem{Nagamatsu} J. Nagamatsu, N. Nakagawa, T. Muranaka, Y. Zenitani and J. Akimitsu: Nature (London) {\bf 410} (2001) 63.  
\bibitem{Kurumaev} Z. Kurmaev, I.I. Lyakhovskaya, J. Kortus, N. Miyata, M. Demeter, M. Neumann, M. Yanagihara, M. Watanabe, T. Muranaka and J. Akimitsu: Cond-mat/0103487. 
\bibitem{Callcott} T.A. Callcott, L. Lin, G.T. Woods, G.P. Zhang, J.R.Thompson, M. Paranthaman and D.L. Ederer: Phys. Rev. B {\bf 64} (2001) 132504.  
\bibitem{Nakamura} J. Nakamura, N. Yamada, K. Kuroki, T.A. Callcott, D.L. Ederer, J.D. Denlinger and R.C.C. Perera: Phys. Rev. B {\bf 64} (2001) 174504. 
\bibitem{Imada} M. Imada: J. Phys. Soc. Jpn. {\bf 70} (2001) 1218. 
\bibitem{Yamaji} K. Yamaji: J. Phys. Soc. Jpn. {\bf 70} (2001) 1476. 
\bibitem{Kortus} J. Kortus, I.I. Mazin, K.D. Belashchenko, V.P. Antropov and L.L. Boyer: Phys. Rev. Lett. {\bf 86} (2001) 4656. 
\bibitem{Belashchenko} K. D. Belashchenko, M. van Schilfgaarde and V. Antrov: Phys. Rev. B {\bf 64} (2001) 092503. 
\bibitem{Satta} G. Satta, G. Profeta, F. Bernardini, A. Continenza and S. Massidda: Phys. Rev. B {\bf 64} (2001) 104507.  
\bibitem{An} J.M. An and W. E. Pickett: Phys. Rev. Lett. {\bf 86} (2001) 4366.
\bibitem{Serrato} A. Reyes-Serrato and D.H. Galv\`an: Cond-mat/0103477. 
\bibitem{Suzuki} S. Suzuki, S. Higai and K. Nakao: J. Phys. Soc. Jpn. {\bf 70} (2001) 1206.  
\bibitem{Rosner} H. Rosner, W.E. Pickett, S.-L. Drechsler, A. Handstein, G. Behr, G. Fuchs, K. Nenkov, K.-H. M\"uller and H. Eschrig: Phys. Rev. {\bf 64} (2001) 144516/1-6.    
\bibitem{Medvedeva} N.I. Medvedeva, A.L. Ivanovskii, J.E. Medvedeva and A.J. Freeman: Phys. Rev. B {\bf 64} (2001) 020502.  
\bibitem{Oguchi} T. Oguchi: unpublished.
\bibitem{Watanabe} M. Watanabe, A. Toyoshima, J. Adachi and A. Yagishita: Nucl. Inst. Meth. Phys. Res. A {\bf 467--468} (2001) 512.
\bibitem{Harada} Y. Harada, H. Ishii, M. Fujisawa, Y. Tezuka, S. Shin, M. Watanabe, Y. Kitajima and A. Yagishita: J. Synchrotron Rad. {\bf 5} (1998) 1013.
\bibitem{Nishibori}  E. Nishibori, M. Takata, M. Sakata, H. Tanaka, T. Muranaka and J. Akimitsu: J. Phys. Soc. Jpn. {\bf 70} (2001) 2252.  
\bibitem{Lee}  S. Lee, H. Mori, T. Masui, Y. Eltsev, A. Yamamoto and S. Tajima: J. Phys. Soc. Jpn. {\bf 70} (2001) 2255.  
\end{thebibliography}
\end{document}